\documentclass[english,pra,aps,amsmath,amssymb,showpacs,preprint]{revtex4}
\usepackage[T1]{fontenc}
\usepackage[latin9]{inputenc}
\usepackage{amsmath}
\usepackage{amssymb}
\usepackage{esint}

\makeatletter
\@ifundefined{textcolor}{}
{%
 \definecolor{BLACK}{gray}{0}
 \definecolor{WHITE}{gray}{1}
 \definecolor{RED}{rgb}{1,0,0}
 \definecolor{GREEN}{rgb}{0,1,0}
 \definecolor{BLUE}{rgb}{0,0,1}
 \definecolor{CYAN}{cmyk}{1,0,0,0}
 \definecolor{MAGENTA}{cmyk}{0,1,0,0}
 \definecolor{YELLOW}{cmyk}{0,0,1,0}
}

\usepackage{subfigure}\usepackage{hyperref}\usepackage{float}\hypersetup{
unicode=false,
pdftoolbar=true,
pdfmenubar=true,
pdffitwindow=false,
pdfstartview={FitH},
pdftitle={Mytitle},
pdfauthor={Author},
pdfsubject={Subject},
pdfcreator={Creator},
pdfproducer={Producer},
pdfkeywords={keywords},
pdfnewwindow=true,
colorlinks=true,
linkcolor=red,
citecolor=blue,
filecolor=magenta,
urlcolor=cyan
}

\makeatletter\newcommand{\Rmnum}[1]{\expandafter\@slowromancap\romannumeral #1@}\makeatother

\makeatother

\usepackage{babel}

\makeatother

\usepackage{babel}

\makeatother

\usepackage{babel}

\makeatother

\usepackage{babel}

\makeatother

\usepackage{babel}

\makeatother

\usepackage{babel}

\makeatother

\usepackage{babel}

\makeatother

\usepackage{babel}

\makeatother

\usepackage{babel}

\makeatother

\usepackage{babel}

\makeatother

\usepackage{babel}
\begin{document}

\title{The geometric and topological interpretation of Berry phase on a
torus}

\author{Da-Bao Yang}

\email{bobydbcn@163.com}

\affiliation{Department of fundamental physics, School of physical science and
technology, Tianjin Polytechnic University, Tianjin 300387, People's
Republic of China}

\author{Kun Meng}

\affiliation{Department of applied physics, School of physical science and technology,
Tianjin Polytechnic University, Tianjin 300387, People's Republic
of China}

\author{Yi-Zhi Wu}

\affiliation{Department of applied physics, School of physical science and technology,
Tianjin Polytechnic University, Tianjin 300387, People's Republic
of China}

\author{Yun-Ge Meng}

\affiliation{Department of fundamental physics, School of physical science and
technology, Tianjin Polytechnic University, Tianjin 300387, People's
Republic of China}

\date{\today}
\begin{abstract}
Illustration of the geometric and topological properties of Berry
phase is often in an obscure and abstract language of fiber bundles.
In this article, we demonstrate these properties with a lucid and
concrete system whose parameter space is a torus. The instantaneous
eigenstate is regarded as a basis. And the corresponding connection
and curvature are calculated respectively. Furthermore, we find the
magnitude of curvature is exactly the Gaussian curvature, which shows
its local property. The topological property is reflected by the integral
over the torus due to Gauss-Bonnet theorem. When we study the property
of parallel transportation of a vector over a loop, we make a conclusion
that the Berry phase is just the angle between the final and initial
vectors. And we also illuminate the geometric meaning of gauge transformation,
which is just a rotation of basis.
\end{abstract}


\maketitle

\section{Introduction}

\label{sec:introduction}

The study of Berry's phase was first introduced by his famous paper
in the context of the cyclic adiabatic evolution and nondegenerate
case \cite{berry1984quantal}. And its connection with fiber bundles
was built by Simon \cite{simon1983holonomy}. It was generalized
to degenerate case by Wilczek and Zee \cite{wilczed1984gauge} as
well as the nonadiabatic case by Aharonov and Anandan \cite{aharonov1987phase}.
And it was also generalized in the mixed state under unitary evolution
by Sj$\ddot{o}$qvist et. al. \cite{sjoqvist2000geometric}. The elaboration
about this field is covered by two monographs \cite{bohm2003geometric,chruscinski2012geometric}.
Moreover many mathematical skills of differential geometry used in
this field could be find in the book written by Flanders \cite{flanders1963differential}.

At the same time, the topological phenomenon in integer quantum Hall
effect was explained by Thouless et. al. \cite{thouless1982quantized},
who proposed the famous TKNN invariant. Haldane predicted that quantum
Hall effect can also be realized in graphene under the background
of applied staggered magnetic field, which broke time reversal symmetry
\cite{haldance1988model}. Nearly two decades later, taking account
of spin-orbit coupling, Kane and Mele began to study a quantum spin
Hall insulator in the context of two layers graphene without breaking
time reversal symmetry \cite{kane2005z2,kane2005quantum}. However,
the intrinsic spin orbital coupling is too tiny to be observed in
experiments. Soon, Bernevig, Hughes and Zhang proposed that the spin
Hall effect might be realized in HgTe/CdTe quantum wells \cite{bernevig2006quantum},
which was verified by K$\ddot{o}$nig et. al. \cite{kronig2007quantum}.
There are also some excellent reviewed articles \cite{hasan2010topological,qi2011topological}
and monograph \cite{bernevig2013topological} .

In 2006, Qi, Wu and Zhang proposed a square lattice as a simple model
of topological insulator, whose Brillouin zone is a two dimensional
torus \cite{qi2006topological}. So the geometric and topological
properties whose base is torus should be elucidated in detail. It
is organised as follows. In the next section, Berry's phases will
be reviewed together with connection and curvature. In Sec. III, in
order to illustrate properties vividly, we will treat a basis as
the instantaneous eigenstate. And the corresponding connection and
curvature will be calculated respectively. Furthermore, we will also
study the topological property. In Sec. IV, the analogy between Berry
phase and parallel transportation will be drawn. And we also elucidate
the geometric meaning of gauge transformation as well. Finally, a
conclusion is drawn in the last section.

\section{Review of Berry's phases}

\label{sec:reviews}

Suppose there exists a quantum system under periodic evolution whose
Hamiltonian depends on some slowly varying parameters, which can be
described as
\[
H(\boldsymbol{R}(t))=H(R_{1}(t),R_{2}(t),\cdots R_{n}(t)),
\]
with $\boldsymbol{R}(T)=\boldsymbol{R}(0)$, where $T$ is the period.
Its evolution is described by Schr$\ddot{o}$dinger equation, which
is
\begin{equation}
i\hbar\frac{d}{dt}|\varPsi(t)\rangle=H(\boldsymbol{R}(t))|\varPsi(t)\rangle.\label{eq:SchrodingerEquation}
\end{equation}
Under adiabatic approximation, the solution to the above equation
can have this form
\begin{equation}
|\varPsi(t)\rangle=c(t)|n,\boldsymbol{R}(t)\rangle,\label{eq:AdiabaticAssumption}
\end{equation}
where $|n,\boldsymbol{R}(t)\rangle$ is an instantaneous eigenstate
of $H(\boldsymbol{R}(t))$, which reads
\begin{equation}
H(\boldsymbol{R}(t))|n,\boldsymbol{R}(t)\rangle=E_{n}|n,\boldsymbol{R}(t)\rangle.\label{eq:InstantaneousEquation}
\end{equation}
The evolution of this system will trace a curve in the parameter space
$\boldsymbol{R}(t)$. Moreover, if the system undergoes periodic motion,
it will trace a closed curve. Suppose the energy eigenstate is nondegenerate,
by substituting Eq. \eqref{eq:AdiabaticAssumption} into Eq. \eqref{eq:SchrodingerEquation}
and multiplying $\langle n,\boldsymbol{R}(t)|$ on the left hand side
at both sides of the equation, one can obtain
\[
|\varPsi(t)\rangle=e^{i\delta}e^{i\gamma}|n,\boldsymbol{R}(t)\rangle
\]
where $\delta$ is called dynamical phase
\[
\delta=-\frac{1}{\hbar}\int_{0}^{T}E(\boldsymbol{R}(t))dt
\]
and $\gamma$ is called Berry's phase or geometric phase \cite{berry1984quantal,bohm2003geometric}
\begin{equation}
\gamma=i\ointop\langle n,\boldsymbol{R}(t)|\frac{\partial}{\partial\boldsymbol{R}}|n,\boldsymbol{R}(t)\rangle\cdot d\boldsymbol{R}.\label{eq:BerryPhase}
\end{equation}
The integrand is called the vector potential, which is
\begin{equation}
\boldsymbol{A}=i\langle n,\boldsymbol{R}(t)|\frac{\partial}{\partial\boldsymbol{R}}|n,\boldsymbol{R}(t)\rangle.\label{eq:BerryVectorPotential}
\end{equation}
And it can be also written as differential one form
\begin{equation}
A=\boldsymbol{A}\cdot d\boldsymbol{R}=i\langle n,\boldsymbol{R}(t)|d|n,\boldsymbol{R}(t)\rangle\label{eq:BerryConnectionOneForm}
\end{equation}
By use of Stokes's theorem, the loop integral can be converted to
a surface integral,
\[
\gamma=\ointop_{c=\partial s}A=\intop_{s}dA.
\]
So we can obtain Berry's curvature
\[
F=dA,
\]
which can also be expressed in vector notation,
\[
\boldsymbol{F}=\boldsymbol{\nabla}\times\boldsymbol{A}.
\]
In the following section, we will elucidate the geometric meaning
of vector potential and Berry phase in a concrete parameter space,
i.e. torus.

\section{Berry's connection, curvature and topology}

\label{sec:Torus}

At first, let's build the parametric equations of torus step by step.
Let's draw a circle on $x-z$ plane whose center is at $(b,0)$ and
radius is $a$, where $b>a$. Its equation can be written as
\[
\boldsymbol{r}=(b+a\cos\psi)\boldsymbol{i}+a\sin\psi\boldsymbol{k},
\]
where $\boldsymbol{i}$, $\boldsymbol{j}$ and $\boldsymbol{k}$ are
unite vectors about $x$, $y$ and $z$ axis respectively. And the
corresponding parametric equations read
\[
\begin{cases}
x= & (b+a\cos\psi)\\
z= & a\sin\psi
\end{cases}
\]
When we rotate this circle along $z-$axis by $SO(2)$ transformation,
a torus can be obtained, whose equation can be expressed as
\[
\left(\begin{array}{c}
x\\
y\\
z
\end{array}\right)=\left(\begin{array}{ccc}
\cos\theta & -\sin\theta & 0\\
\sin\theta & \cos\theta & 0\\
0 & 0 & 1
\end{array}\right)\left(\begin{array}{c}
b+a\cos\psi\\
0\\
a\sin\psi
\end{array}\right).
\]
Thus, the final equation of torus can be written as \cite{edvards2002calculus}
\[
\begin{cases}
x= & (b+a\cos\psi)\cos\theta\\
y= & (b+a\cos\psi)\sin\theta\\
z= & a\sin\psi
\end{cases}
\]
 For convenient, we want to transform it to a vector notation, which
is
\begin{equation}
\boldsymbol{r}=x(\theta,\psi)\boldsymbol{i}+y(\theta,\psi)\boldsymbol{j}+z(\theta,\psi)\boldsymbol{k}.\label{eq:position}
\end{equation}
For physical realization, we may contrive a spin subject to a magnetic
field, whose Hamiltonian takes this form
\[
H=-\boldsymbol{\mu}\cdot\boldsymbol{B}
\]
where $\boldsymbol{\mu}$ is the magnetic moment and $\boldsymbol{B}$
is the magnetic field. And the components of $\boldsymbol{B}$ are
\[
\begin{cases}
B_{x}= & (\beta+\alpha\cos\psi)\cos\theta\\
B_{y}= & (\beta+\alpha\cos\psi)\sin\theta\\
B_{z}= & \alpha\sin\psi,
\end{cases}
\]
where $\beta>\alpha$. We are not intend to calculate its connection
nor curvature in this article, but to uncover their underlying geometric
meaning. At first, We can differentiate $\boldsymbol{r}(\theta,\psi)$
\eqref{eq:position} to obtain the basis vector of tangent plane of
torus, which is
\begin{equation}
d\boldsymbol{r}=\frac{\partial\boldsymbol{r}}{\partial\theta}d\theta+\frac{\partial\boldsymbol{r}}{\partial\psi}d\psi.\label{eq:displacement}
\end{equation}
By an explicit calculation, we can obtain
\[
\boldsymbol{\varepsilon_{\theta}}=\frac{\partial\boldsymbol{r}}{\partial\theta}=-(b+a\cos\psi)\sin\theta\boldsymbol{i}+(b+a\cos\psi)\cos\theta\boldsymbol{j}
\]
and
\[
\boldsymbol{\varepsilon_{\psi}}=\frac{\partial\boldsymbol{r}}{\partial\psi}=-a\sin\psi\cos\theta\boldsymbol{i}-a\sin\psi\sin\theta\boldsymbol{j}+a\cos\psi\boldsymbol{k}.
\]
And the lengths of $\boldsymbol{\varepsilon_{\theta}}$ and $\boldsymbol{\varepsilon_{\psi}}$
are
\[
h_{\theta}=b+a\cos\psi
\]
and
\[
h_{\psi}=a
\]
respectively. From the above equations, the orthonormal basis vectors
of the tangent plane can be get after an observation,
\begin{equation}
\boldsymbol{e_{\theta}}=-\sin\theta\boldsymbol{i}+\cos\theta\boldsymbol{j}\label{eq:basistheta}
\end{equation}
and
\begin{equation}
\boldsymbol{e_{\psi}=}-\sin\psi\cos\theta\boldsymbol{i}-\sin\psi\sin\theta\boldsymbol{j}+\cos\psi\boldsymbol{k}\label{eq:basisphi}
\end{equation}
So by substituting Eq. \eqref{eq:basistheta} and \eqref{eq:basisphi}
into Eq. \eqref{eq:displacement}, we can obtain
\[
d\boldsymbol{r}=(b+a\cos\psi)d\theta\boldsymbol{e_{\theta}}+ad\psi\boldsymbol{e_{\psi}}.
\]
Furthermore, the area element on the torus can be calculated by multiplying
the two coefficients before $\boldsymbol{e_{\theta}}$ and $\boldsymbol{e_{\psi}}$
directly, which is
\begin{equation}
d\sigma=a(b+a\cos\psi)d\theta d\psi.\label{eq:AreaElementOfTorus}
\end{equation}
In order to comparison with instantaneous eigenstate \eqref{eq:InstantaneousEquation},
we define a new orthogonal basis on the surface of torus, which takes
the following form \cite{chruscinski2012geometric}
\begin{equation}
\boldsymbol{e_{+}}=\frac{1}{\sqrt{2}}(\boldsymbol{e_{\theta}}+i\boldsymbol{e_{\psi}})\label{eq:basisplus}
\end{equation}
and
\begin{equation}
\boldsymbol{e_{-}}=\frac{1}{\sqrt{2}}(\boldsymbol{e_{\theta}}-i\boldsymbol{e_{\psi}}).\label{eq:basisminus}
\end{equation}
The connection 1-form can be evaluated easily as
\begin{equation}
A=i\boldsymbol{e_{+}}^{*}\cdot d\boldsymbol{e_{+}}=\sin\psi d\theta,\label{eq:ConnectionOneForm}
\end{equation}
which has a similar form as Berry's connection one form \eqref{eq:BerryConnectionOneForm}.
And it can be translated into the language of vector potential, which
is
\begin{equation}
\boldsymbol{A}=\frac{\sin\psi}{b+a\cos\theta}\boldsymbol{e_{\theta}}\label{eq:vectorpotential}
\end{equation}
Next, we calculate the curvature two-form
\[
F=dA=-\cos\psi d\theta\wedge d\psi,
\]
which can also has vector form
\[
\boldsymbol{F}=-\frac{\cos\psi}{a(b+a\cos\psi)}\boldsymbol{e_{\bot},}
\]
Where $\boldsymbol{e_{\bot}}$ is the normal vector of the torus. Next,
we want to calculate Gaussian curvature of torus, which will uncover
the geometric meaning of Berry's curvature. At first, the normal vector
of torus is evaluated by

\[
\begin{array}{ccl}
\boldsymbol{e_{\bot}} & = & \boldsymbol{e_{\theta}}\times\boldsymbol{e_{\psi}}\\
 & = & \cos\theta\cos\psi\boldsymbol{i}+\sin\theta\cos\psi\boldsymbol{j}+\sin\psi\boldsymbol{k}.
\end{array}
\]
In order to obtain the corresponding area element of surface of a
unit sphere, we differentiate $\boldsymbol{e_{\bot}}$ with respect
to $\theta$ and $\psi$ respectively, which is
\[
d\boldsymbol{e_{\bot}}=\cos\psi d\theta\boldsymbol{e_{\theta}}+d\psi\boldsymbol{e_{\psi}.}
\]
Hence, the area element on the unit sphere is obtained by multiplying
the two coefficients before $\boldsymbol{e_{\theta}}$ and $\boldsymbol{e_{\psi}}$
directly, which reads
\begin{equation}
d\sigma^{\prime}=\cos\psi d\theta d\psi.\label{eq:AreaElementOfSphere}
\end{equation}
By use of Eq. \eqref{eq:AreaElementOfTorus} and Eq. \eqref{eq:AreaElementOfSphere},
the Gaussian curvature can be calculated as
\[
K=\frac{d\sigma^{\prime}}{d\sigma}=\frac{\cos\psi}{a(b+a\cos\theta)},
\]
which is exactly the magnitude of Berry's curvature $\boldsymbol{F}$.
So Berry's curvature reflects the local geometric property of torus.

The direct calculation of integral of Gaussian curvature over the
torus reflects its topological property, which is
\[
\varoiint Kd\sigma=-\varoiintop F=0.
\]
It is certainly in accordance with Gauss-Bonnet theorem, which takes
this form
\[
\varoiint Kd\sigma=4\pi(1-g),
\]
where $g$ is called genus, which is equal to the number of holes
of the manifold. In the case of torus, $g=1$.

\section{Parallel transport and Gauge transformation}

\label{sec:ParallelTransportAndGaugeTransformation}

Suppose there exists a tangential vector on the torus, which takes
this form
\[
\boldsymbol{v}=v_{+}\boldsymbol{e_{+}}+v_{-}\boldsymbol{e_{-}}.
\]
Moreover, the above vector can also represented in the original basis
as
\[
\boldsymbol{v}=v_{\theta}\boldsymbol{e_{\theta}}+v_{\psi}\boldsymbol{e_{\psi}}.
\]
Due to the basis transformation between them, i.e. Eq. \eqref{eq:basisplus}
and \eqref{eq:basisminus}, the transformations between components
can also be built, which are
\begin{equation}
v_{+}=\frac{1}{\sqrt{2}}(v_{\theta}-iv_{\psi})\label{eq:TransformationOfComponents}
\end{equation}
and
\[
v_{-}=\frac{1}{\sqrt{2}}(v_{\theta}+iv_{\psi})
\]
This vector will be transported parallel along a curve on the surface
of torus. Due to the parallel transportation, it is natural to require
the vector is invariant, which is
\[
d\boldsymbol{v}=0.
\]
By substituting Eq. \eqref{eq:basisplus} and Eq. \eqref{eq:basisminus}
into the above equation, we can obtain
\[
(dv_{+}-i\sin\psi v_{+}d\theta)\boldsymbol{e_{+}}+(dv_{-}+i\sin\psi v_{-}d\theta)\boldsymbol{e_{-}}=0,
\]
where the normal component is discarded in accordance with the requirement
of tangential vector. Because $\boldsymbol{e_{+}}$ and $\boldsymbol{e_{-}}$
are orthonormal basis, the following results can be get directly,
which are
\begin{equation}
dv_{+}-i\sin\psi v_{+}d\theta=0\label{eq:paralelltransportation}
\end{equation}
and
\[
dv_{-}+i\sin\psi v_{-}d\theta=0.
\]
In comparison with Berry phase, let's imagine our vector is along
the $\boldsymbol{e_{+}}$ axis initially, hence the component along
$\boldsymbol{e_{-}}$ vanishes. So the vector takes the following
form
\[
\boldsymbol{v_{0}}=v_{0}\boldsymbol{e_{+}}.
\]
The parallel transportation along a closed curve can be described
by the Eq. \eqref{eq:paralelltransportation}, so this equation can
be solved as
\[
v_{+}=v_{0}\exp\left(i\oint\sin\psi d\theta\right).
\]
By substituting the vector potential \eqref{eq:vectorpotential} into
the above equation, it can be converted to
\begin{equation}
v_{+}=v_{0}\exp\left(i\oint\boldsymbol{A}\cdot d\boldsymbol{r}\right),\label{eq:holonomy}
\end{equation}
where $d\boldsymbol{r}=(b+a\cos\psi)d\theta\boldsymbol{e_{\theta}}+ad\psi\boldsymbol{e_{\psi}}$
is the infinitesimal displacement vector along the surface of this
torus. From the above Eq. \eqref{eq:holonomy}, we can make a conclusion
that the direction of the final vector maybe different from its original
one, where this phenomenon is called holonomy in the realm of differential
geometry. However, when $\psi=0$ or $\psi=\pi$, which are two geodesics
on the torus, the loop integral in Eq. \eqref{eq:holonomy} vanishes.
This is due to the fact that when the vector is transported parallel
along a geodesic, the direction will not change after a closed contour.
In addition, we would illuminate the geometric feature of the loop
integral that is equivalent to Berry phase. For simplicity, we denote
\begin{equation}
\gamma=\oint\boldsymbol{A}\cdot d\boldsymbol{r}.\label{eq:gamma}
\end{equation}
Substituting Eq. \eqref{eq:TransformationOfComponents} and Eq. \eqref{eq:gamma}
into Eq. \eqref{eq:holonomy},
\[
\frac{1}{\sqrt{2}}(v_{+\theta}-iv_{+\psi})=\frac{1}{\sqrt{2}}(v_{0\theta}-iv_{0\psi})(\cos\gamma+i\sin\gamma).
\]
After comparing the real and imaginary parts of the above equation,
one can obtain that
\[
\begin{pmatrix}v_{+\theta}\\
v_{+\psi}
\end{pmatrix}=\begin{pmatrix}\cos\gamma & \sin\gamma\\
-\sin\gamma & \cos\gamma
\end{pmatrix}\begin{pmatrix}v_{0\theta}\\
v_{0\psi}
\end{pmatrix},
\]
which means that the vector is rotated clockwise by angle $\gamma$
after parallel transportation along a closed curve.

Next, we want to study the geometric meaning of gauge transformation
\cite{berry1984quantal,bohm2003geometric,chruscinski2012geometric}.
First of all, let's do a gauge transformation on $\boldsymbol{e_{+}}$
\eqref{eq:basisplus}, which is
\begin{equation}
\boldsymbol{e_{+}^{\prime}}=\exp(-i\chi)\boldsymbol{e_{+}},\label{eq:GaugeTransformationOnBasisPlus}
\end{equation}
where in general $\chi$ is a function of $\theta$ and $\phi$. By
plugging the above Eq. \eqref{eq:GaugeTransformationOnBasisPlus}
into connection \eqref{eq:ConnectionOneForm}, we can obtain the new
connection which takes the following form
\[
\begin{array}{ccl}
A^{\prime} & = & i\boldsymbol{e_{+}^{\prime}}^{*}\cdot d\boldsymbol{e_{+}^{\prime}}\\
 & = & A+d\chi,
\end{array}
\]
which is exactly the same as the gauge transformation of Berry's connection.
In order to illustrate the geometric meaning of gauge transformation,
we go back to original basis $\boldsymbol{e_{\theta}}$and $\boldsymbol{e_{\phi}}$.
By substituting Eq. \eqref{eq:basistheta} and Eq. \eqref{eq:basisphi}
into Eq. \eqref{eq:GaugeTransformationOnBasisPlus} and comparing
the real and imaginary parts at the both side of equations, we can
obtain
\[
\begin{pmatrix}\boldsymbol{e_{\theta}^{\prime}} & \boldsymbol{e_{\psi}^{\prime}}\end{pmatrix}=\begin{pmatrix}\boldsymbol{e_{\theta}} & \boldsymbol{e_{\psi}}\end{pmatrix}\begin{pmatrix}\cos\chi & \sin\chi\\
-\sin\chi & \cos\chi
\end{pmatrix},
\]
 is just the basis transformation. The primed basis vectors are obtained
by rotating the unprimed ones counterclockwise by angle $\chi$.

\section{Conclusions and Acknowledgements }

\label{sec:ConclusionAndAccknowledgements}

In this paper, we illustrate the geometric and topological property
of Berry phase whose parameter space is a torus. The instantaneous
eigenstate is like the basis $\boldsymbol{e_{+}}$ \eqref{eq:basisplus}
and $\boldsymbol{e_{-}}$ \eqref{eq:basisminus}. And the corresponding
connection and curvature are calculated respectively. Furthermore,
we find the magnitude of curvature is exactly the Gaussian curvature,
which demonstrate its local property. The topological property is
reflected by the integral over the torus due to Gauss-Bonnet theorem.
When we study the property of parallel transportation of a vector
over a loop, we make a conclusion that the Berry phase is just the
angle between the final and initial vectors. And we also elucidate
the geometric meaning of gauge transformation, which is just a rotation
of basis.

D.B.Y. thanks his wife and mother for taking care of the infant. D.B.Y.
is supported by NSF ( Natural Science Foundation ) of China under
Grant No. 11447196. K. M. is supported by NSF of China under grant
No.11447153. And Y.Z.W. is supported by Scientific Research Plan Project
of Tianjin Municipal Education Commission (No. 2017KJ097) .

\end{document}